\newcommand{\diracslash}[1]{#1\llap{/\kern2pt}}

\newcommand{\be}{\begin{equation}}
\newcommand{\ee}{\end{equation}}
\newcommand{\bea}{\begin{eqnarray}}
\newcommand{\eea}{\end{eqnarray}}
\newcommand{\ba}[1]{\begin{array}{#1}}
\newcommand{\ea}{\end{array}}

\documentclass[twocolumn,superscriptaddress,secnumarabic,amssymb,nobibnotes,nofootinbib,aps,prl,showpacs]{revtex4}
\usepackage{graphicx}
\begin{document}
\setlength{\topmargin}{-0.05in}

\title{Diagnosis of human breast cancer through wavelet transform of polarized fluorescence}

\author{Nrusingh C. Biswal}
 \affiliation{Indian Institute of Technology, Kanpur-208 016, India}

\author{Sharad Gupta}
 \affiliation{Indian Institute of Technology, Kanpur-208 016, India}

\author{Asima Pradhan}
\affiliation{Indian Institute of Technology, Kanpur-208 016,
India}

\author{Prasanta K. Panigrahi}
 \affiliation{
Physical Research Laboratory, Navrangpura, Ahmedabad 380 009,
India}
 \affiliation{
School of Physics, University of Hyderabad, Hyderabad-500 046,
India}

\date{\today}

\def\be{\begin{equation}}
\def\ee{\end{equation}}
\def\bearr{\begin{eqnarray}}
\def\eearr{\end{eqnarray}}
\def\zbf#1{{\bf {#1}}}
\def\bfm#1{\mbox{\boldmath $#1$}}
\def\hf{\frac{1}{2}}

\begin{abstract}
Wavelet transform of polarized fluorescence spectroscopic data of
human breast tissues is found to reliably differentiate normal and
malignant tissue types and isolate characteristic biochemical
signatures of cancerous tissues, which can possibly be used for
diagnostic purpose. A number of parameters capturing spectral
variations and subtle changes in the diseased tissues in the
visible wavelength regime are clearly identifiable in the wavelet
domain. These investigations, corroborated with tissue phantoms,
indicate that the observed differences between malignant tumor and
normal samples are primarily ascribable to the changes in
concentration of porphyrin and density of cellular organelles
present in tumors.
\end{abstract}


\maketitle

Breast cancer has emerged as the most common disease amongst women
\cite{Wang}. Although the risk factor for Asian women has been
estimated to be one-fifth to one-tenth that of women in North
America and Western Europe, it still is the second most malignant
condition \cite{Lippman,Pesto}. Apart from genetic predisposition,
a number of factors like diet, exercise, environment, etc., are
being recognized to play major roles in the growth of the disease
\cite{Lahad}. Early diagnosis is still not possible through
conventional diagnostic techniques. If diagnosed early, breast
cancer is also one of the most treatable forms of cancer. The
requirement of continuous monitoring for breast malignancy of a
significant percentage of women population has led to an intense
search for safe, reliable and fast diagnostic methods.

Optical diagnosis techniques are now emerging as viable tools for
tumor detection. Of these, fluorescence techniques are being
increasingly employed to investigate both morphological and
biochemical changes in different tissue types, for eventual
application in the detection of tumors at an early stage
\cite{RecRef}. Fluorescence spectroscopy is well suited for the
diagnosis of cancerous tissues because of its sensitivity to
minute variations in the amount and the local environment of the
native fluorophores present in the tissues
\cite{Alfano,Tang1,Tang2,Kortum,Wilson}. Morphological changes
prevalent in tumors, such as enlargement and hyperchromasia of
nuclei, overcrowding and irregular cellular arrangement are known
to alter light propagation and scattering properties in such media
and hence affect the fluorescence spectra \cite{Backman}. A number
of fluorophores ranging from structural proteins to various
enzymes and coenzymes, some of which participate in the cellular
oxidation-reduction processes, are present in the human tissue and
can be excited by ultraviolet and visible light \cite{Alfano}. The
fluorophores, FAD (Flavin Adenine Dinucleotide), its derivatives
and porphyrins are particularly useful as fluorescent markers,
since they fluoresce in the higher wavelength visible region, when
excited by lower wavelength visible light, thereby avoiding the
potentially harmful ultraviolet radiation.

The fluorescence emission can differ significantly in normal and
cancerous tissues due to the differences in concentrations of
absorbers \cite{Feld,NRam} and scatterers, as also the scatterer
sizes \cite{Perel}. The absorption in the visible range occurs
primarily due to the presence of blood, whose amounts vary in
various tissue types \cite{NRam1}. The presence of scatterers
leads to randomization of light, thereby generating a depolarized
component in the fluorescence spectra. Polarized fluorescence
spectroscopy is useful in isolating the characteristic spectral
features from the diffuse background. The parallel component of
the fluorescence suffers fewer scattering events. In comparison,
the intensity of the perpendicular component is not only affected
more by scatterers, but is also quite sensitive to absorption,
since the path traversed by the same in the tissue medium is more.
Hence, the difference of parallel and perpendicular intensities,
apart from being relatively free from the diffusive component
\cite{Backman1}, can be quite sensitive to microscopic biochemical
changes including the effects of absorption in different tissue
types.

A number of studies conducted so far have established certain
broad morphological and biochemical changes occurring in tumor
tissues, which leave characteristic signatures in the spectral
domain \cite{NRam2}. The analyses of spectral data involve both
physical \cite{Gard,Wu,Durkin,Laxmi} and statistical
\cite{NRam1,Dillon} modelling of tissue types, as also statistical
methods, e.g., principal component analysis for extracting
distinguishing parameters for diagnostic purposes \cite{NRam1} .
The fact that biological tissues are complex systems, possessing
substantial variations among individual patients, depending upon
various factors such as age, progress of the disease, etc., makes
modelling of the same rather difficult. In using statistical
tools, difficulty often arises in relating the statistically
significant quantities to physically transparent spectral
variables. In recent times, wavelet transform has emerged as a
powerful tool for the analysis of transient data and is
particularly useful in disentangling characteristic variations at
different scales \cite{Daube}. This linear transform isolates
local features and leads to a convenient dimensional reduction of
the data in the form of low-pass (average) coefficients,
resembling the data itself. The wavelet or high-pass coefficients,
at various levels, encapsulate the variations at corresponding
scales. The higher-level coefficients, particularly the global
parameters associated with them, like power, are less contaminated
by statistical and experimental uncertainties present in the data.
An earlier study, of the perpendicular component of the
fluorescence spectra, by some of the present authors has indicated
the usefulness of wavelet transform in identifying characteristic
spectral features \cite{Nidhi}.

Here, we present the results of a systematic analysis of the
wavelet transform of the fluorescence spectra from human breast
tissues for malignant and normal tissues. The difference between
parallel and perpendicular components of the fluorescence spectra
is subjected to this analysis, since the same is comparatively
free of the diffusive component. A number of parameters, capturing
spectral variations and subtle changes in the intensity profile of
the diseased tissues, as compared to their normal counterparts,
are identified in the wavelet domain. Based on earlier
investigations and the present study of tissue phantoms, the
physical origin of these distinguishing parameters can be
primarily ascribed to the changes in the concentration of
porphyrins and the density of cellular organelles present in
tumors \cite{Perel,Nair}.

In total, 28 breast cancer tissue samples were studied; out of
these, 23 samples came with their normal counterparts. The tissue
samples were excited by 488nm wavelength polarized light and the
parallel and perpendicularly polarized fluorescence light were
measured from 500 to 700 nm. Differences of parallel and
perpendicular components of fluorescence intensity ($I_\parallel -
I_\perp$) versus wavelength profiles for all the tissue samples
were analyzed by Haar wavelets \cite{Chui}.

We have identified three independent parameters, derived from the
coefficients in the transform domain, which differentiate cancer
and normal tissues quite accurately. The first parameter is the
local maxima in the third quarter of the fourth level low-pass
coefficients. As will be elaborated later, this feature owes its
origin to porphyrin emission \cite{NRam2,Nair}. The other two
parameters are based on wavelet high-pass coefficients,
representing both global and characteristic local variations of
the fluorescence spectra. In the domain of these three parameters,
all the malignant and normal tissues studied here could be
accurately differentiated. The five unpaired samples were used as
checks for the consistency of two of the chosen parameters, since
one of the parameters is a ratio, which involves both tissue
types. Studies on tissue phantoms, corroborating the above choice
of parameters and the inferences about the aforementioned
biochemical changes in the tissues are presented below for
comparison.

In the fourth level low-pass coefficients, the one originating
from the fluorescence signals around 630 nm of the original data
is found to be considerably higher in cancer tissues as compared
to the corresponding normal ones. This is possibly due to the
presence of more porphyrin as well as scattering agents. A
particularly noisy fluorescence data of ($I_\parallel - I_\perp$)
from cancer and normal human breast tissues does not reveal
significant differences (Fig.1). However, the low-pass wavelet
coefficients of the same data (Fig.1, inset) capture these
differences quite remarkably, highlighting the usefulness of
wavelet analysis.
\begin{figure}
\includegraphics[width=2.50in]{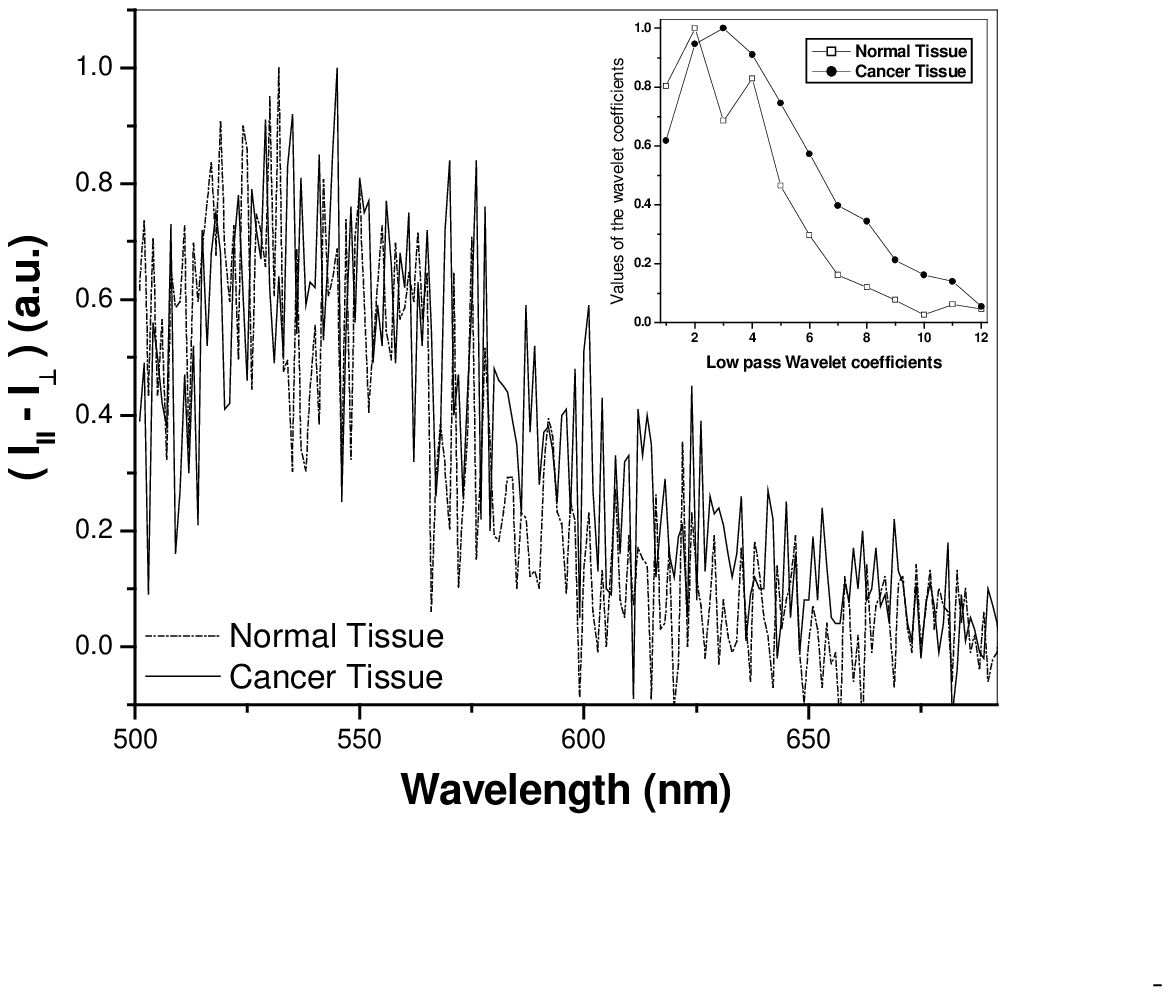}
\caption{Plot of the difference of parallel and perpendicular
components of the fluorescence spectra for tumor and normal
tissues. Inset shows the corresponding fourth level low-pass
coefficients.} \label{figdel}
\end{figure}

The local maxima at third quarter of fourth level low-pass
coefficients of cancer samples are more than 0.1 while those of
normal tissues are less than 0.1, with a sensitivity of $100 \%$
and specificity of $83\%$ (Fig. 2). It should be noted here that
the values for normal tissues which are more than 0.1 still show
lower values than the corresponding tumors, consistent with all
the other samples. Thus intra-patient diagnosis gives a clear
distinction between cancer and normal tissues. Variations in
inter-patient diagnosis may be due to the fact that, the growth of
tumor depends on genetic (major genes, modifier genes) and
non-genetic factors (birth, age, weight/diet, exercise,
environmental exposures, etc) \cite{Lahad}.

\begin{figure}
\includegraphics[width=2.50in]{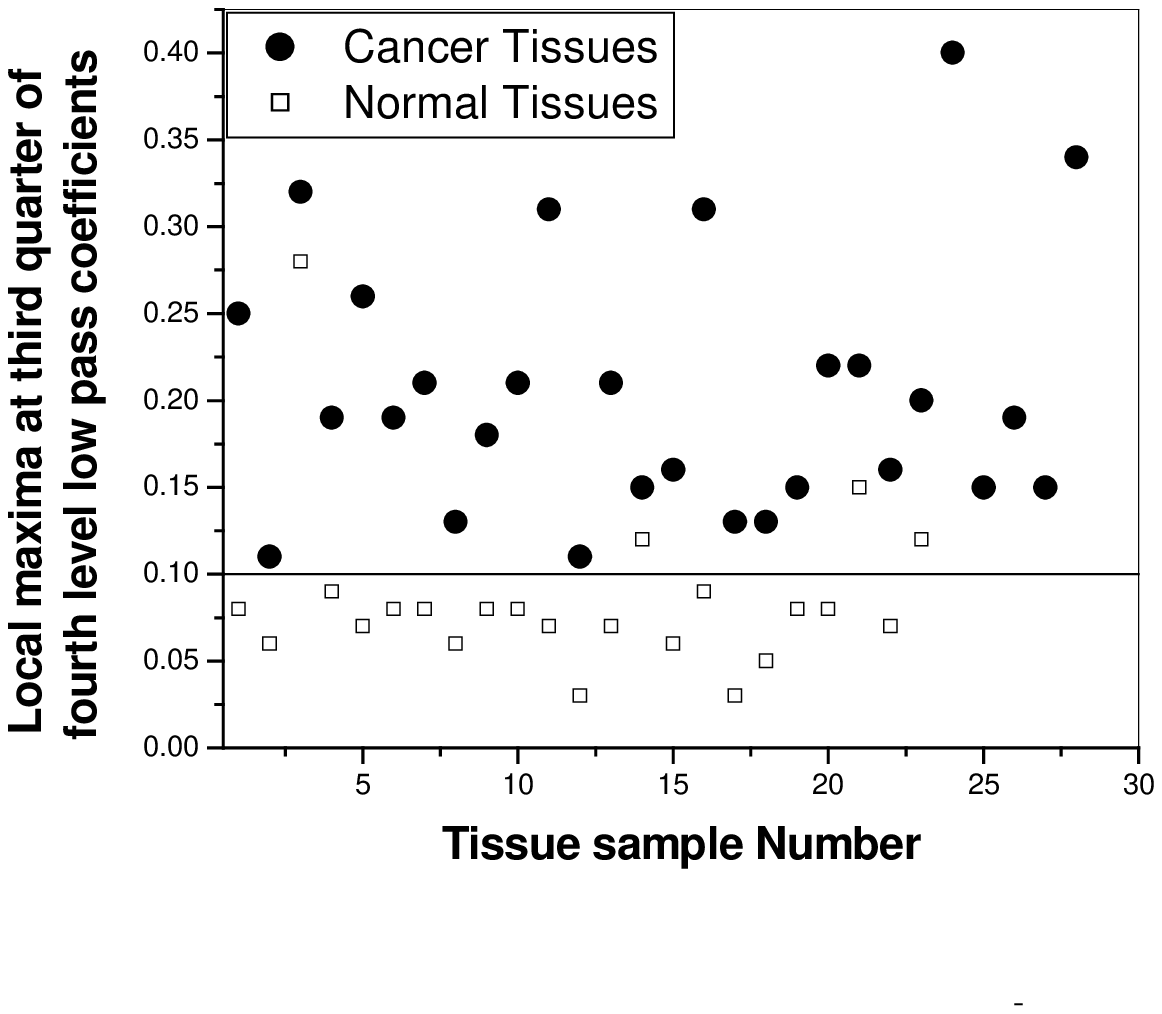}
\caption{. Local maxima at third quarter of fourth level low-pass
coefficients of cancer and normal breast tissues} \label{figdel}
\end{figure}
 An important observation here is that the 630 nm band gets emphasized only in the
fourth level low-pass coefficients. This band is masked by other
noisy signals at the third level and is averaged out at the fifth
level (Fig. 3). In cases where the fourth level does not highlight
this band, the previous level does.

\begin{figure}
\includegraphics[width=2.50in]{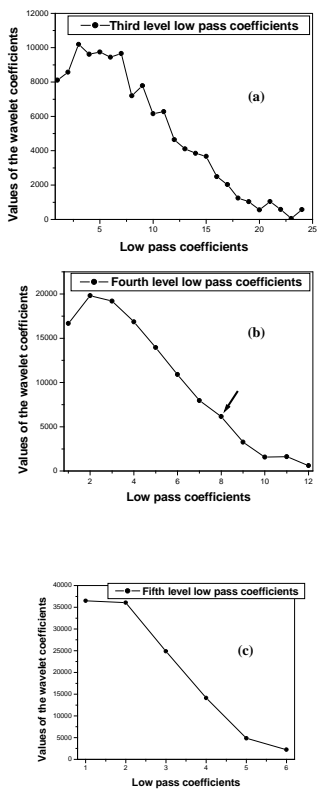}
\caption{Low-pass wavelet coefficients of a cancer tissue spectra,
(a) third level, (b) fourth level, and (c) fifth level. The
fourth-level coefficients highlight the 630 nm weak emission
peak.} \label{figdel}
\end{figure}

FAD and porphyrins are the major fluorophores that fluoresce in
the visible wavelength regime, with peak intensities at 530 and
630 nm respectively. These fluorophores are considered as contrast
agents for cancer detection \cite{Alfano,Wilson,NRam2}. It has
been suggested that deficiency in ferrochelatase, the enzyme
required for conversion of protoporphyrin IX (PpIX) to heme, in
tumors results in accumulation of PpIX in these tissues relative
to the normal ones (10)\cite{Wilson}. Such accumulation changes
the relative concentration of these fluorophores thus altering the
fluorescence spectra significantly, which in turn changes the peak
heights of the emission bands of the two fluorophores. The
scattering centers are known to enhance the fluorescence intensity
\cite{Biswal}. Thus the large size of cell suspensions, higher
density of cells and accumulation of more porphyrin in tumors all
contribute to a small peak at 630 nm wavelength region.

Studies of tissue models show that the 630 nm band gets enhanced
at the fourth level low-pass coefficients of phantoms, with an
increase in the scatterer concentrations (Fig. 4a) as well as with
increase in porphyrin concentrations (Fig. 4b). A small peak
around 630 nm is clearly visible at suitable concentrations.
Significantly, in these tissue phantoms too, the third and fifth
level low-passes do not highlight the 630 nm band, as observed in
tissue samples.

\begin{figure}
\includegraphics[width=2.50in]{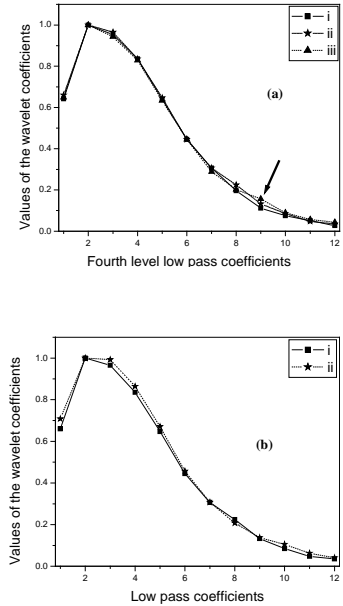}
\caption{(a). Fourth level low-pass coefficients of phantoms with
fixed FAD (20 M) and porphyrin (10 M) concentrations and varying
scatterer concentrations, (i)  s =10 mm-1, (ii)  s =20 mm-1, and
(iii)  s =30 mm-1. Inset shows the values of the 7th to 10th
low-pass wavelet coefficients of the same phantoms, highlighting
the variations of a local peak. (b) Fourth level low-pass
coefficients of phantoms, with fixed FAD (20 M) and scatterer (10
M) concentrations and varying porphyrin concentrations (i) 10 M,
and (ii) 20 M.} \label{figdel}
\end{figure}
The power spectra at different levels are defined as the sum of
the square of high-pass coefficients at those levels.
Normalization of the power spectra is done by dividing it by the
sum of the square of intensities at all the wavelengths. In twenty
two paired samples, it was found that the tumors have lower power
at the third level as compared to their normal counterparts with a
sensitivity of $96\%$.

It was also found that, in case of the cancer tissues, the third
wavelet coefficient at the fifth level (originating from the
fluorescence emission at 580 to 596 nm region in the original
spectrum), is less negative than those of the normal ones. This
implies that the normal tissue fluorescence spectra fall more
sharply than those of the cancer tissues. Out of 28 cancer
samples, which includes 23 paired and 5 unpaired tissues, 21
samples have third coefficients less than -0.31; out of 23 normal
tissue samples, 14 samples have third coefficients more than
-0.31. However, intra-patient diagnosis by high-pass coefficients
shows that the third coefficient, for 17 out of 23 paired samples
of normal tissues, is more than that of cancerous ones. Hence, for
this coefficient, the cancer to normal ratio is less than one,
with a sensitivity of $74\%$.

It may be noted that the above-mentioned three parameters also
distinguish tumors of different grades. It is found that for
grades I and II cancerous tissues the values of the local low-pass
maxima at the third quadrant are less than 0.2, but more than 0.2
in the grade III cancers, with a sensitivity of $75\%$.  At third
level, the power ratio is less than 0.3 for grade I and grade II
cancers and is between 0.3 to 0.8 for grade III cancers.

In conclusion, the systematic separation of variations at
different wavelength scales and the broad spectral features in the
wavelet domain pinpoints several quantifiable parameters to
distinguish cancer and normal tissues. These distinguishable
features are related with the biochemical and morphological
changes, as is also evident from the phantom study. The fact that
these characteristic signatures are based on higher level wavelet
coefficients make them robust and less susceptible to experimental
and statistical uncertainties. The need for the early
identification and constant monitoring of breast cancer for a
large population makes this method eminently suitable since the
same can be automated.

\end{document}